\documentclass[aps,pra,reprint,amsmath,amssymb,superscriptaddress,nofootinbib,longbibliography]{revtex4-2}

\usepackage{graphicx}
\usepackage[utf8]{inputenc}
\usepackage[T1]{fontenc}
\usepackage{xcolor}
\usepackage{soul}
\usepackage{amssymb}
\usepackage{bm}

\begin{document}

\title{Magnetocaloric performance of RE$_{11}$Co$_4$In$_9$ (RE = Tb, Er)}

\author{Stanis\l{}aw Baran}
\email{stanislaw.baran@uj.edu.pl}
\affiliation{M.~Smoluchowski Institute of Physics, Jagiellonian University,
prof. Stanis\l{}awa \L{}ojasiewicza 11, PL-30-348 Krak\'ow, Poland}
\author{Altifani Rizky Hayyu}
\affiliation{M.~Smoluchowski Institute of Physics, Jagiellonian University,
prof. Stanis\l{}awa \L{}ojasiewicza 11, PL-30-348 Krak\'ow, Poland}
\author{Yuriy Tyvanchuk}
\affiliation{Department of Analytical Chemistry, Ivan Franko National
University of Lviv, Kyryla i Mefodiya 6, UA-79005 Lviv, Ukraine}
\author{Andrzej Szytu\l{}a}
\affiliation{M.~Smoluchowski Institute of Physics, Jagiellonian University,
prof. Stanis\l{}awa \L{}ojasiewicza 11, PL-30-348 Krak\'ow, Poland}

\date{\today}

\begin{abstract}

The magnetocaloric effect in RE$_{11}$Co$_4$In$_9$ (RE = Tb, Er) has been studied by means of
magnetometric measurements in the function of temperature and applied magnetic field. The maximum
magnetic entropy change ($-\Delta S_{\mathrm{M}}^{\mathrm{max}}$) at magnetic flux density change
($\Delta\mu_{0}H$) 0-9~T has been determined to be 5.51~J$\cdot$kg$^{-1}\cdot$K$^{-1}$ at 47.4~K for Tb$_{11}$Co$_4$In$_9$
and 14.28~J$\cdot$kg$^{-1}\cdot$K$^{-1}$ at 12.3~K for Er$_{11}$Co$_4$In$_9$, while temperature averaged entropy change
(TEC) with 3~K span equals 5.50 and 14.14~J$\cdot$kg$^{-1}\cdot$K$^{-1}$ for RE = Tb and Er, respectively. The
relative cooling power (RCP) and refrigerant capacity (RC) equal respectively 522.1 and 391.2~J$\cdot$kg$^{-1}$ in
Tb$_{11}$Co$_4$In$_9$ and 605.2 and 463.1~J$\cdot$kg$^{-1}$ in Er$_{11}$Co$_4$In$_9$.

\bigskip

\noindent \textbf{keywords}: rare earth intermetallics, magnetocaloric effect, magnetic entropy change, relative cooling power, refrigerant capacity

\end{abstract}

\maketitle

\section{Introduction}
\label{intro}

Magnetocaloric effect (MCE) is nowadays of great scientific interest due to its application in
magnetic refrigeration (MR)~\cite{franco2018magnetocaloric_effect}. Recently, promising magnetocaloric properties have been reported
for RE$_{11}$Co$_4$In$_9$ (RE = Gd, Dy, Ho)~\cite{zhang2020structural}.

The RE$_{11}$Co$_4$In$_9$ (RE = Gd--Er) intermetallics are known to crystallize in an orthorhombic
crystal structure of the Nd$_{11}$Pd$_4$In$_9$-type (space group $Cmmm$)~\cite{tyvanchuk2012R11Co4In9,zhang2020structural,baran2021crystal}.
Within the $ab$-plane, the structure consists of the RECo$_2$ (AlB$_2$-type) and REIn (CsCl-type) fragments in the 9:2 ratio~\cite{sojka2008Nd11Pd4In9}.
Along the $c$-axis, the layers composed of the rare earth atoms ($z = 0$) are separated by layers containing the Co and In atoms $\left(z = \frac{1}{2}\right)$.
In this complex crystal structure the rare earth atoms occupy five nonequivalent Wyckoff sites.

The measurements of magnetization vs temperature, undertaken at magnetic flux density $\mu_0H$ of 2~T (20~kOe),
show a transition from para- to ferromagnetic state at 86, 37 and 20~K for RE = Gd, Dy and Ho, respectively~\cite{zhang2020structural}.
The more recent measurements, performed at a low magnetic flux density of 0.005~T (50~Oe), reveal a cascade of magnetic transitions at
95, 85, 70 and 35~K (RE = Tb), 88, 28 and 19~K (RE = Dy), 33 and 10~K (RE = Ho) with only exception for RE = Er, where a single transition,
characteristic of formation of an antiferromagnetic order below 5.4~K, is observed~\cite{baran2021crystal}. The effective magnetic
moments in RE$_{11}$Co$_4$In$_9$ (RE = Gd--Er) are close to the values expected for the free RE$^{3+}$ ions, indicating that magnetism
in the investigated compounds is related to the rare earth magnetic moments while the Co atoms remain either non-magnetic or their
magnetic moments are tiny and negligible when compared to high rare earth moments. Isothermal magnetization curves do not show
saturation for RE~=~Gd, Dy and Ho at $T = 3$~K and $\mu_0H = 7$~T (70~kOe)~\cite{zhang2020structural} as well as for RE~=~Tb, Dy, Ho and Er
at $T = 2$~K and $\mu_0H = 9$~T (90~kOe)~\cite{baran2021crystal}. It is worth noting that the primary isothermal magnetization curve for
RE~=~Er, collected at 2~K, shows a metamagnetic transition at 0.06~T (0.6~kOe), indicating that the low-temperature antiferromagnetic order
can be turned into a ferromagnetic one by application of relatively low magnetic field~\cite{baran2021crystal}.

The magnetocaloric properties have been reported only for RE = Gd, Dy and Ho~\cite{zhang2020structural}. In the vicinity of the
respective Curie temperature, the magnetic entropy change reaches 10.95, 4.66 and 12.29~J$\cdot$kg$^{-1}\cdot$K$^{-1}$ for RE~=~Gd, Dy and Ho,
respectively. The temperature averaged entropy change (TEC) with 3~K span shows the values of 10.93 (RE~=~Gd), 4.64 (Dy) and
12.09~J$\cdot$kg$^{-1}\cdot$K$^{-1}$ (Ho). Relative cooling power (RCP) and refrigerant capacity (RC) are found to be 538.1 and 405.9~J$\cdot$kg$^{-1}$
(RE~=~Gd), 213.9 and 165.9~J$\cdot$kg$^{-1}$ (Dy) and 475.2 and 357.4~J$\cdot$kg$^{-1}$ (Ho).

In this work, the magnetocaloric properties of Tb$_{11}$Co$_4$In$_9$ and Er$_{11}$Co$_4$In$_9$ are reported, as derived from measurements
of magnetization in the function of temperature and applied magnetic field (up to 9~T). These results extend the knowledge on magnetocaloric
properties to all members of the RE$_{11}$Co$_4$In$_9$ (RE = Gd--Er) family of compounds.

\section{Materials and methods}

The samples investigated in this work are the same samples as those reported in the previous study~\cite{baran2021crystal}. The reader
interested in synthesis procedure and crystal structure parameters can find all this information in Ref.~\cite{baran2021crystal}.

The magnetocaloric effect in RE$_{11}$Co$_4$In$_9$ (RE = Tb, Er) has been investigated with the use of a vibrating sample
magnetometer (VSM) option of the Physical Properties Measurement System by Quantum Design, equipped with a superconducting magnet running
up to 9~T. In order to collect the data, the following procedure has been performed:
\begin{itemize}

\item Demagnetization in the paramagnetic state by decreasing oscillating magnetic field.

\item Cooling down to 2~K.

\item Setting the desired value of magnetic flux density (the data were collected at 0.5, 1.0, ..., 9.0~T).

\item Collecting magnetization vs temperature data up to 200~K (Tb$_{11}$Co$_4$In$_9$) or 70~K (Er$_{11}$Co$_4$In$_9$).

\item Repetition of the above steps for the next value of magnetic flux density.

\end{itemize}

\section{Results and discussion}

\begin{figure}[!ht]
\begin{center}
\includegraphics[bb=0 2 504 630,width=0.48\textwidth]
        {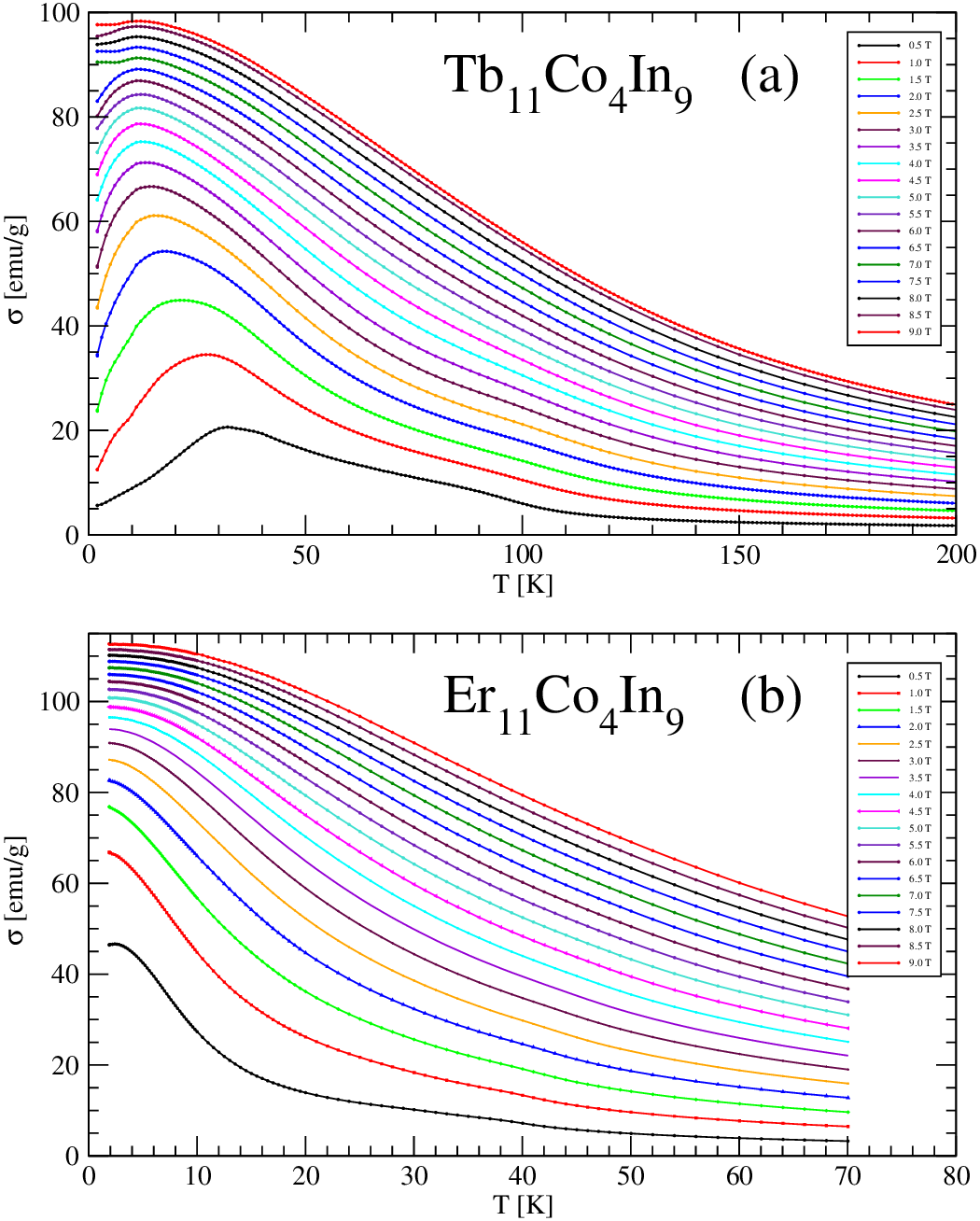}
\end{center}
\caption{
Magnetization vs temperature curves for (a) Tb$_{11}$Co$_4$In$_9$ and (b) Er$_{11}$Co$_4$In$_9$.
The curves were collected at selected values of magnetic flux density ranging from 0.5 to 9.0~T.}
\label{fig:M_vs_T}
\end{figure}

Fig.~\ref{fig:M_vs_T} shows the magnetization vs temperature curves collected for RE$_{11}$Co$_4$In$_9$ (RE = Tb, Er)
at selected values of magnetic flux density. All the curves have an inflection point characteristic of the transition
from para- to ferro-/ferrimagnetic state. A maximum visible in the $\sigma(T)$ curves for Tb$_{11}$Co$_4$In$_9$
indicates the development of an antiferromagnetic contribution to the magnetic structure at low temperatures. With
the increase of the applied magnetic field, the maximum shifts to lower temperatures and finally disappears, indicating
the suppression of the antiferromagnetic component by applying a high magnetic field.

\begin{figure}[!ht]
\begin{center}
\includegraphics[bb=8 10 553 702,width=0.48\textwidth]
        {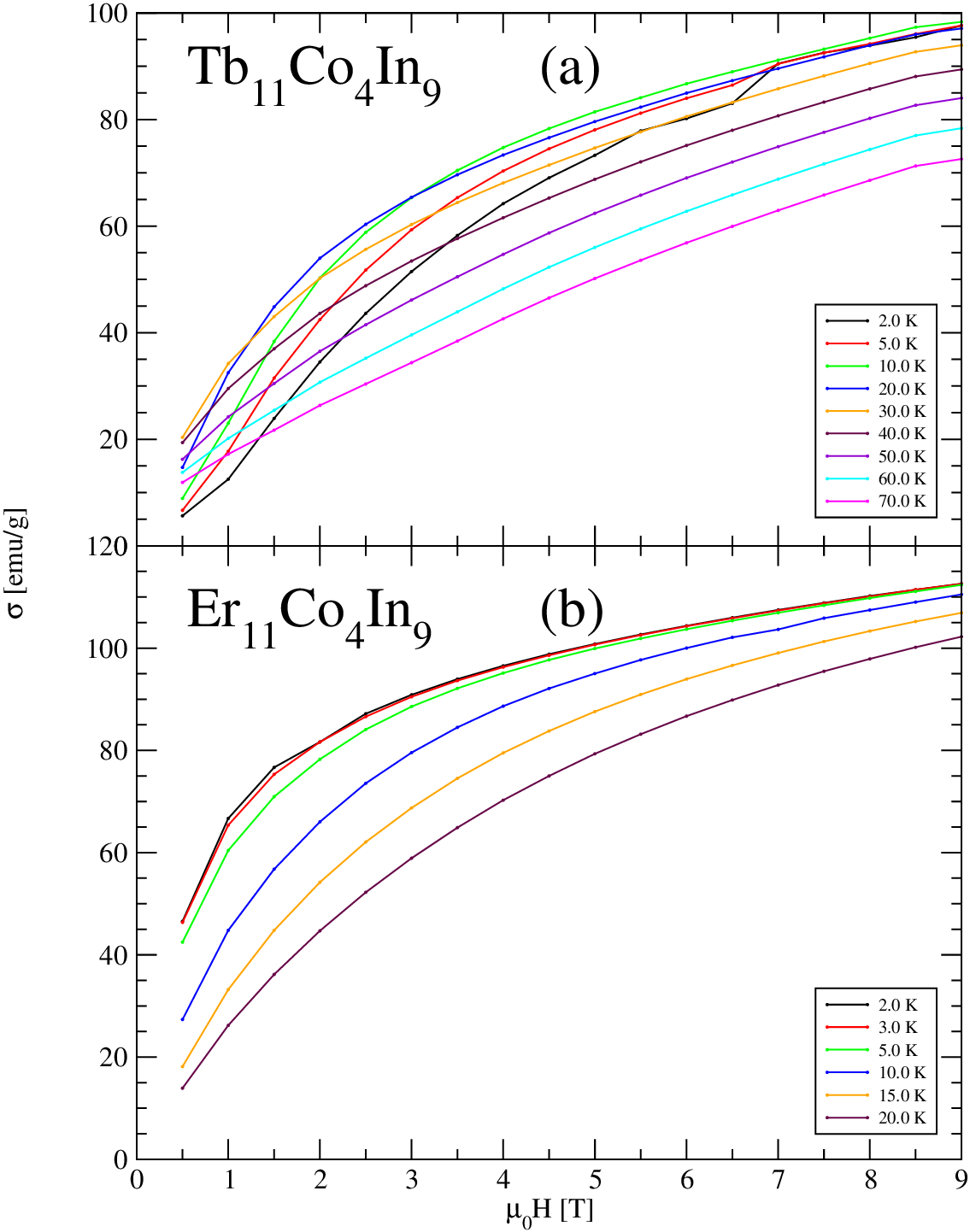}
\end{center}
\caption{
Magnetization vs magnetic flux density curves for (a) Tb$_{11}$Co$_4$In$_9$ and (b) Er$_{11}$Co$_4$In$_9$,
as calculated from the raw data shown in Fig.~\ref{fig:M_vs_T}.}
\label{fig:M_vs_H}
\end{figure}

This result is in agreement with the magnetization vs magnetic flux density curves presented in Fig.~\ref{fig:M_vs_H}.
The curves for Tb$_{11}$Co$_4$In$_9$, collected at the lowest temperatures, show a distinct metamagnetic transition
indicating that antiferromagnetic contribution to the magnetic order is turned into the ferro/ferrimagnetic one by
application of the external magnetic field. The metamagnetic transition disappears with increasing temperature.

\begin{figure}[!ht]
\begin{center}
\includegraphics[bb=2 10 550 715,width=0.48\textwidth]
        {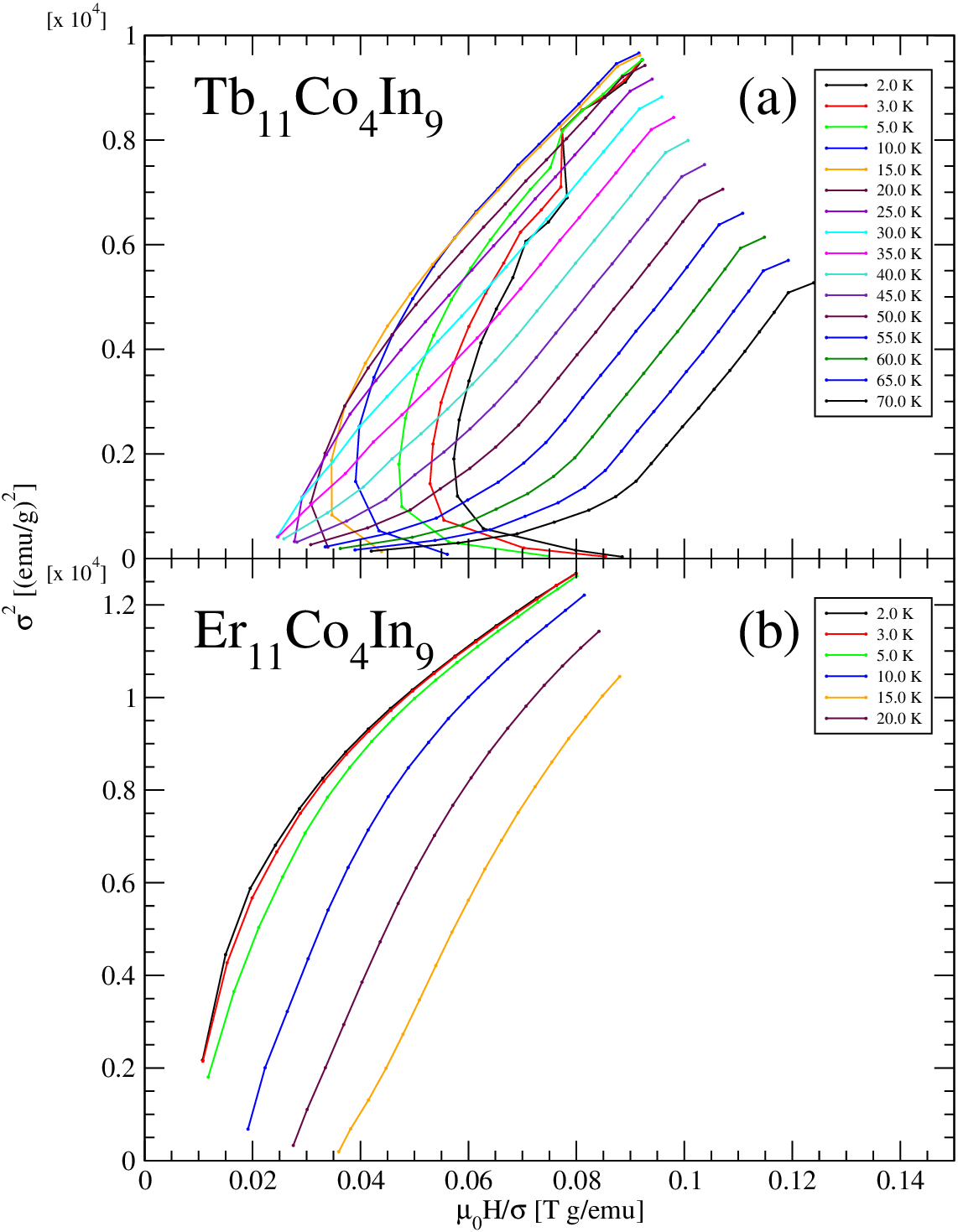}
\end{center}
\caption{
The $\sigma^2(\mu_0 H/\sigma)$ Arrott curves at selected temperatures for (a) Tb$_{11}$Co$_4$In$_9$ and (b) Er$_{11}$Co$_4$In$_9$.}
\label{fig:M2_vs_H_over_M}
\end{figure}

In order to ensure that the magnetic transition corresponding to the maximum magnetic entropy change is the second-order phase 
transition (SOPT), not the first-order phase transition (FOPT), and therefore no thermal hysteresis of
magnetic entropy change is expected, the Arrott curves at selected temperatures have been calculated (see Fig.~\ref{fig:M2_vs_H_over_M}).
The theory predicts that the sign of the slope of Arrott curve corresponds to the character of the magnetic phase transition, i.e. it is
positive for SOPT and negative for FOPT~\cite{banerjee1964generalised_approach}. It is clearly visible in Fig.~\ref{fig:M2_vs_H_over_M}
that all curves for Er$_{11}$Co$_4$In$_9$ have positive slope and only some low-temperature curves for Tb$_{11}$Co$_4$In$_9$
have negative slope within a limited range. Therefore, the FOPT in Tb$_{11}$Co$_4$In$_9$ is related to the appearance of the
low-temperature antiferromagnetic contribution to the magnetic order. It has to be mentioned that all Arrot curves in the vicinity
of temperatures at which magnetic entropy change reaches its maximum (i.e. 47.4~K in Tb$_{11}$Co$_4$In$_9$ and 12.3~K in Er$_{11}$Co$_4$In$_9$
for magnetic flux density change 0-9~T) have positive slope, indicating that the corresponding magnetic phase transition is SOFT.

\begin{figure}[!ht]
\begin{center}
\includegraphics[bb=-1 2 504 630,width=0.48\textwidth]
        {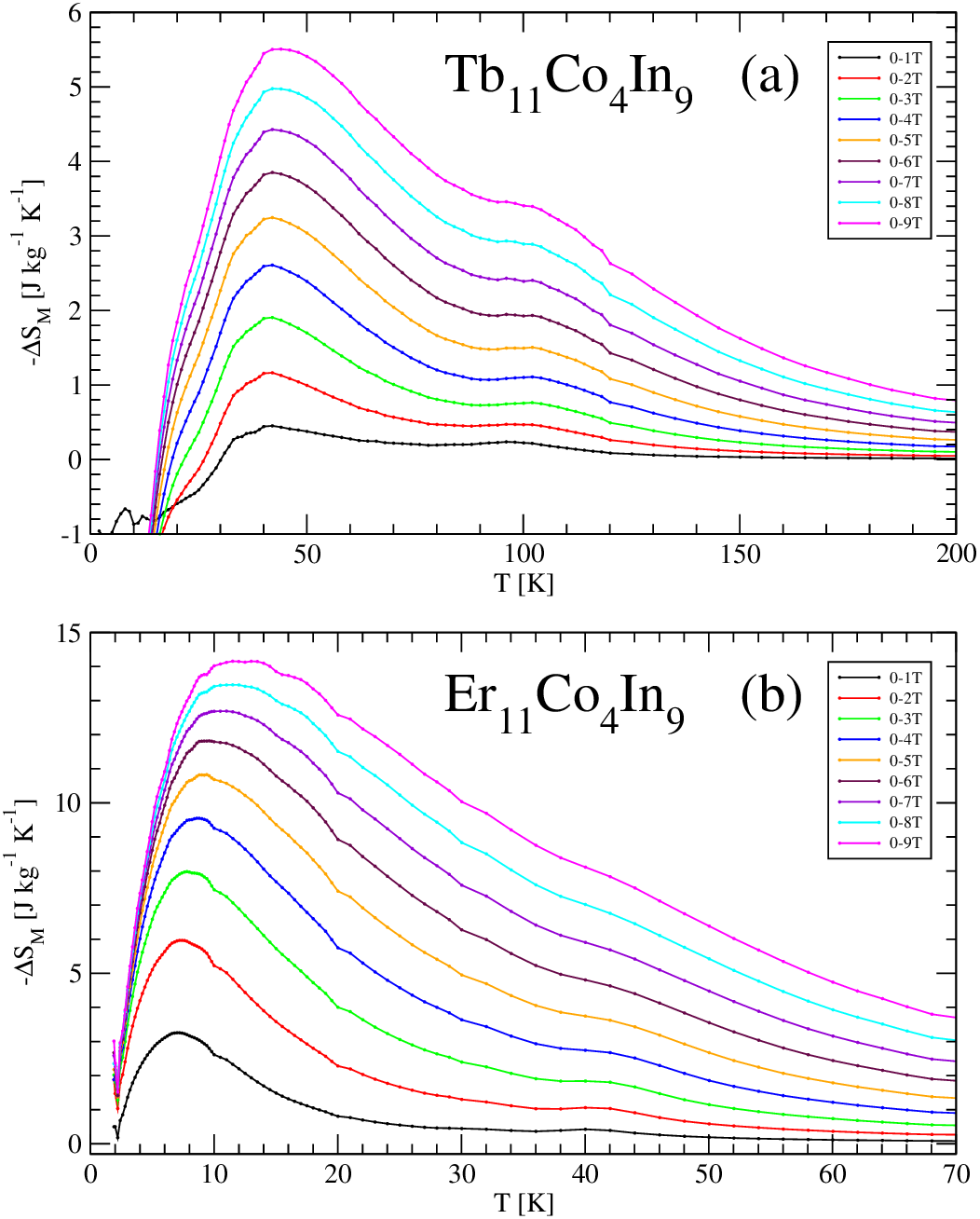}
\end{center}
\caption{
Magnetic entropy change vs temperature at selected values of the magnetic flux density change for (a) Tb$_{11}$Co$_4$In$_9$ and (b)
Er$_{11}$Co$_4$In$_9$.}
\label{fig:magn_entr_vs_T}
\end{figure}

In order to calculate the magnetic entropy change under isothermal conditions, the following well-known equation has been used:

\begin{equation}
\Delta S_{\mathrm{M}}(T, \Delta\mu_{0}H) = \int_{\mu_{0}H_i}^{\mu_{0}H_f} \left( \frac{\partial\sigma(T,\mu_{0}H)}{\partial T} \right)_{\mu_{0}H}  \,\mathrm{d}\mu_{0}H
\label{eqn:magn_entr_chng}
\end{equation}

\noindent where $\Delta\mu_{0}H$ is a change of the magnetic flux density, defined as a difference between the final ($\mu_{0}H_f$) and initial ($\mu_{0}H_i$)
flux densities, while $\left( \frac{\partial\sigma(T,\mu_{0}H)}{\partial T} \right)_{\mu_{0}H}$ is a derivative of magnetization over temperature at fixed
magnetic flux density $\mu_{0}H$~\cite{tishin2003magnetocaloric}. As it is commonly assumed to report magnetic entropy change with respect to the initial
magnetic flux density equal to zero ($\mu_{0}H_i=0$), the same convention is used in the current study. Fig.~\ref{fig:magn_entr_vs_T} shows magnetic entropy
change in function of temperature for selected values of the magnetic flux density change. The maximum magnetic entropy change for $\Delta\mu_{0}H$ = 0-9~T
reaches 5.51~J$\cdot$kg$^{-1}\cdot$K$^{-1}$ at 47.4~K for Tb$_{11}$Co$_4$In$_9$ and 14.28~J$\cdot$kg$^{-1}\cdot$K$^{-1}$ at 12.3~K for Er$_{11}$Co$_4$In$_9$. The values of maximum
magnetic entropy change for different magnetic flux density changes between 0-1 up to 0-9~T are listed in Table~\ref{tbl:tec}.

\begin{figure}[!ht]
\begin{center}
\includegraphics[bb=4 3 499 720,width=0.48\textwidth]
        {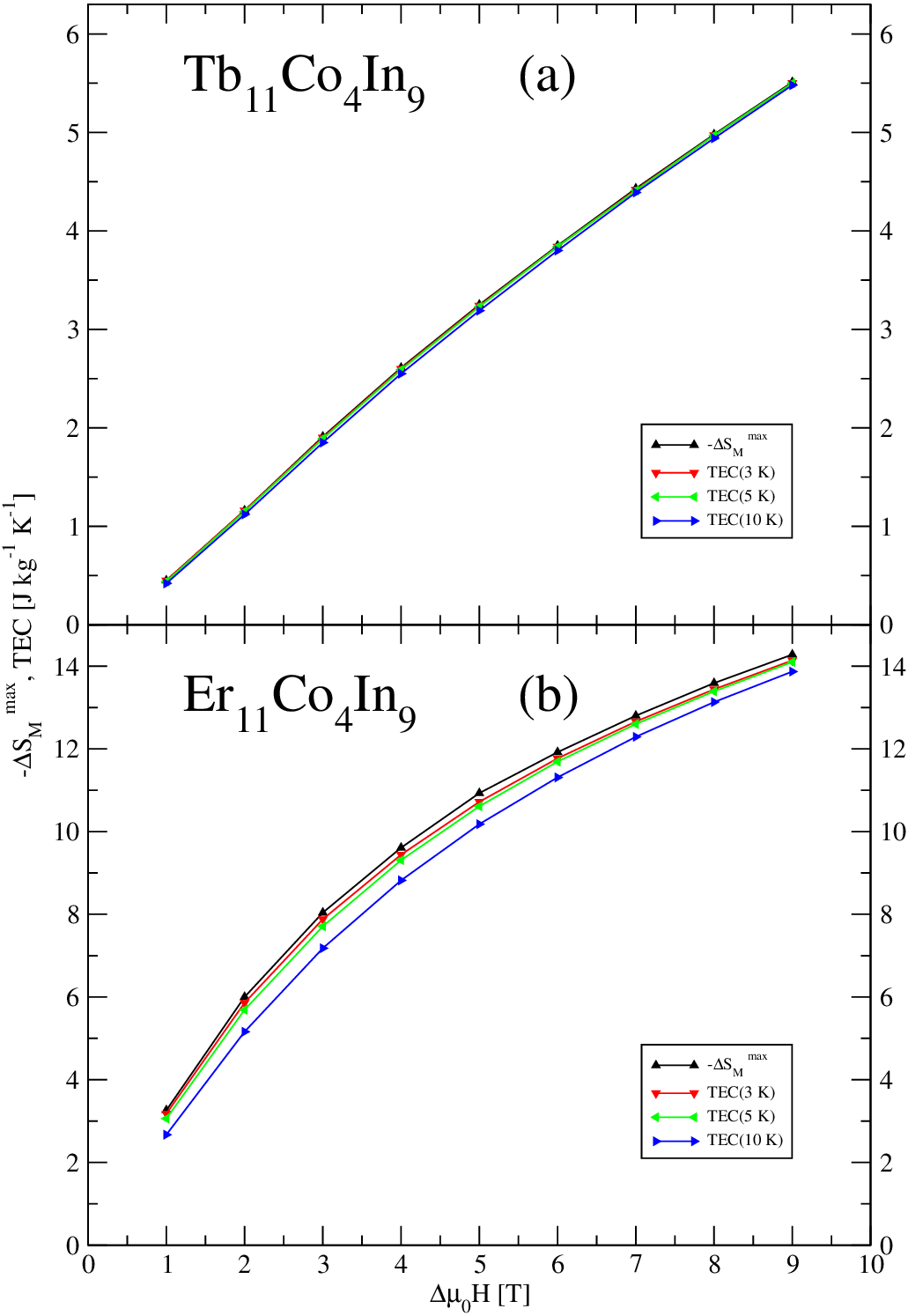}
\end{center}
\caption{
Maximum magnetic entropy change ($-\Delta S_{\mathrm{M}}^{\mathrm{max}}$) together with temperature averaged magnetic entropy
change ($TEC$) over 3, 5 and 10~K in function of magnetic flux density change $\Delta\mu_{0}H$ for
(a) Tb$_{11}$Co$_4$In$_9$ and (b) Er$_{11}$Co$_4$In$_9$.}
\label{fig:tec}
\end{figure}

Another, recently introduced parameter characterizing a magnetocaloric material is the temperature averaged magnetic
entropy change ($TEC$) over a given temperature span ($\Delta T_\mathrm{lift}$), defined by the following equation~\cite{griffith2018material-based}:

\begin{equation*}
TEC(\Delta T_\mathrm{lift},\Delta\mu_{0}H) =
\end{equation*}
\begin{equation}
\frac{1}{\Delta T_\mathrm{lift}} \max_{T_\mathrm{mid}}
\int_{T_\mathrm{mid}-\frac{\Delta T_\mathrm{lift}}{2}}^{T_\mathrm{mid}+\frac{\Delta T_\mathrm{lift}}{2}} \Delta S_{\mathrm{M}}(T,\Delta\mu_{0}H)\,\mathrm{d}T
\label{eqn:tec}
\end{equation}

\noindent where the integral is maximized with respect to the $T_\mathrm{mid}$ parameter which is a center of the temperature span ($\Delta T_\mathrm{lift}$).
Fig.~\ref{fig:tec} shows maximum magnetic entropy change ($-\Delta S_{\mathrm{M}}^{\mathrm{max}}$) together with temperature averaged magnetic entropy
change ($TEC$) over 3, 5 and 10~K in function of magnetic flux density change $\Delta\mu_{0}H$ for RE$_{11}$Co$_4$In$_9$ (RE = Tb, Er).
The values of $-\Delta S_{\mathrm{M}}^{\mathrm{max}}$ and $TEC$ for the whole RE$_{11}$Co$_4$In$_9$ (RE = Gd-Er) series of intermetallics is listed
in Table~\ref{tbl:tec}. While comparing $-\Delta S_{\mathrm{M}}^{\mathrm{max}}$ and $TEC$ at magnetic flux density change of 0-7~T for compounds
with different chemical compositions, it is well visible that the values of $TEC$(3~K) are very close to those of $-\Delta S_{\mathrm{M}}^{\mathrm{max}}$
and they slightly decrease with increasing temperature span $T_\mathrm{lift}$. However, even for $T_\mathrm{lift}=10$~K, the corresponding temperature
averaged magnetic entropy changes ($TEC$(10~K)) equal at least 94~\% of the respective $-\Delta S_{\mathrm{M}}^{\mathrm{max}}$, indicating that
RE$_{11}$Co$_4$In$_9$ (RE = Gd-Er) show good magnetocaloric performance over wide temperature span.

{
\renewcommand{\arraystretch}{.67}
\begin{table*}[!ht]
\begin{footnotesize}
\caption{The values of $-\Delta S_{\mathrm{M}}^{\mathrm{max}}$, TEC(3~K), TEC(5~K) and TEC(10~K) under various $\Delta\mu_{0}H$
        for RE$_{11}$Co$_4$In$_9$ (RE = Gd--Er). The data taken from Ref.~\cite{zhang2020structural} are
        shown in the $\Delta\mu_{0}H$ range up to 0-7~T while those reported in this work in the range up to 0-9~T.}

{\begin{tabular*}{0.99\textwidth}{l}
\end{tabular*}}

  \begin{flushleft}
  \vspace{-0.5 cm}
  \normalsize Gd$_{11}$Co$_4$In$_9$ (data from Ref.~\cite{zhang2020structural})
  \end{flushleft}
  \vspace{-0.2 cm}
  \footnotesize
  \begin{tabular*}{0.99\textwidth}{@{\extracolsep{\fill}}lllll}
    \hline
    $\Delta\mu_{0}H[\mathrm{T}]$ & $-\Delta S_{\mathrm{M}}^{\mathrm{max}}$ & TEC(3~K) & TEC(5~K) & TEC(10~K)\\
& [J$\cdot$kg$^{-1}\cdot$K$^{-1}$] & [J$\cdot$kg$^{-1}\cdot$K$^{-1}$] & [J$\cdot$kg$^{-1}\cdot$K$^{-1}$] & [J$\cdot$kg$^{-1}\cdot$K$^{-1}$] \\
    \hline
 0-1 & 2.92 & 2.79 & 2.68 & 2.41\\ 
 0-2 & 5.03 & 4.96 & 4.88 & 4.59\\
 0-3 & 6.55 & 6.52 & 6.48 & 6.22\\
 0-4 & 7.86 & 7.83 & 7.79 & 7.55\\
 0-5 & 9.01 & 8.92 & 8.88 & 8.70\\
 0-6 & 10.02 & 10.01 & 9.93 & 9.72\\ 
 0-7 & 10.95 & 10.93 & 10.85 & 10.65\\ 
    \hline
  \end{tabular*}
  \begin{flushleft}
  \normalsize Tb$_{11}$Co$_4$In$_9$ (this work)
  \end{flushleft}
  \vspace{-0.2 cm}

  \footnotesize
  \begin{tabular*}{0.99\textwidth}{@{\extracolsep{\fill}}lllll}
    \hline
    $\Delta\mu_{0}H[\mathrm{T}]$ & $-\Delta S_{\mathrm{M}}^{\mathrm{max}}$ & TEC(3~K) & TEC(5~K) & TEC(10~K)\\
& [J$\cdot$kg$^{-1}\cdot$K$^{-1}$] & [J$\cdot$kg$^{-1}\cdot$K$^{-1}$] & [J$\cdot$kg$^{-1}\cdot$K$^{-1}$] & [J$\cdot$kg$^{-1}\cdot$K$^{-1}$] \\
    \hline
 0-1 & 0.45 & 0.45 & 0.44 & 0.42\\ 
 0-2 & 1.16 & 1.16 & 1.15 & 1.12\\
 0-3 & 1.91 & 1.90 & 1.89 & 1.85\\
 0-4 & 2.61 & 2.60 & 2.59 & 2.55\\
 0-5 & 3.25 & 3.24 & 3.23 & 3.19\\ 
 0-6 & 3.85 & 3.84 & 3.84 & 3.80\\ 
 0-7 & 4.43 & 4.42 & 4.41 & 4.39\\ 
 0-8 & 4.98 & 4.97 & 4.97 & 4.94\\ 
 0-9 & 5.51 & 5.50 & 5.50 & 5.48\\ 
    \hline
  \end{tabular*}

  \begin{flushleft}
  \normalsize Dy$_{11}$Co$_4$In$_9$ (data from Ref.~\cite{zhang2020structural})
  \end{flushleft}
  \vspace{-0.2 cm}
  \footnotesize
  \begin{tabular*}{0.99\textwidth}{@{\extracolsep{\fill}}lllll}
    \hline
    $\Delta\mu_{0}H[\mathrm{T}]$ & $-\Delta S_{\mathrm{M}}^{\mathrm{max}}$ & TEC(3~K) & TEC(5~K) & TEC(10~K)\\
& [J$\cdot$kg$^{-1}\cdot$K$^{-1}$] & [J$\cdot$kg$^{-1}\cdot$K$^{-1}$] & [J$\cdot$kg$^{-1}\cdot$K$^{-1}$] & [J$\cdot$kg$^{-1}\cdot$K$^{-1}$] \\
    \hline
 0-1 & 0.38 & 0.37 & 0.37 & 0.34\\ 
 0-2 & 1.14 & 1.12 & 1.18 & 1.12\\
 0-3 & 2.09 & 2.06 & 2.02 & 1.94\\
 0-4 & 2.88 & 2.84 & 2.80 & 2.71\\
 0-5 & 3.53 & 3.51 & 3.47 & 3.38\\ 
 0-6 & 4.09 & 4.08 & 4.05 & 3.98\\ 
 0-7 & 4.66 & 4.64 & 4.61 & 4.55\\ 
    \hline
  \end{tabular*}

  \begin{flushleft}
  \normalsize Ho$_{11}$Co$_4$In$_9$ (data from Ref.~\cite{zhang2020structural})
  \end{flushleft}
  \vspace{-0.2 cm}
  \footnotesize
  \begin{tabular*}{0.99\textwidth}{@{\extracolsep{\fill}}lllll}
    \hline
    $\Delta\mu_{0}H[\mathrm{T}]$ & $-\Delta S_{\mathrm{M}}^{\mathrm{max}}$ & TEC(3~K) & TEC(5~K) & TEC(10~K)\\
& [J$\cdot$kg$^{-1}\cdot$K$^{-1}$] & [J$\cdot$kg$^{-1}\cdot$K$^{-1}$] & [J$\cdot$kg$^{-1}\cdot$K$^{-1}$] & [J$\cdot$kg$^{-1}\cdot$K$^{-1}$] \\
    \hline
 0-1 & 0.91 & 0.90 & 0.88 & 0.83\\ 
 0-2 & 2.95 & 2.90 & 2.86 & 2.74\\
 0-3 & 5.23 & 5.11 & 5.03 & 4.81\\
 0-4 & 7.34 & 7.17 & 7.06 & 6.78\\
 0-5 & 9.22 & 9.00 & 8.87 & 8.55\\ 
 0-6 & 10.84 & 10.61 & 10.47 & 10.13\\ 
 0-7 & 12.29 & 12.09 & 11.90 & 11.54\\ 
    \hline
  \end{tabular*}

  \begin{flushleft}
  \normalsize Er$_{11}$Co$_4$In$_9$ (this work)
  \end{flushleft}
  \vspace{-0.2 cm}
  \footnotesize
  \begin{tabular*}{0.99\textwidth}{@{\extracolsep{\fill}}lllll}
    \hline
    $\Delta\mu_{0}H[\mathrm{T}]$ & $-\Delta S_{\mathrm{M}}^{\mathrm{max}}$ & TEC(3~K) & TEC(5~K) & TEC(10~K)\\
& [J$\cdot$kg$^{-1}\cdot$K$^{-1}$] & [J$\cdot$kg$^{-1}\cdot$K$^{-1}$] & [J$\cdot$kg$^{-1}\cdot$K$^{-1}$] & [J$\cdot$kg$^{-1}\cdot$K$^{-1}$] \\
    \hline
    0-1 & 3.26 & 3.18 & 3.06 & 2.67\\
    0-2 & 6.00 & 5.86 & 5.69 & 5.16\\
    0-3 & 8.04 & 7.88 & 7.71 & 7.18\\
    0-4 & 9.61 & 9.44 & 9.31 & 8.82\\
    0-5 & 10.93 & 10.72 & 10.61 & 10.18\\
    0-6 & 11.92 & 11.77 & 11.69 & 11.31\\
    0-7 & 12.80 & 12.66 & 12.60 & 12.29\\
    0-8 & 13.59 & 13.44 & 13.39 & 13.13\\
    0-9 & 14.28 & 14.14 & 14.10 & 13.87\\
    \hline
  \end{tabular*}
\label{tbl:tec}
\end{footnotesize}
\end{table*}
}

Apart from the above discussed $-\Delta S_{\mathrm{M}}^{\mathrm{max}}$ and TEC, there are two more parameters that provide useful information about
MCE in a given material -- these are: refrigerant capacity (RC)~\cite{wood_potter1985general_analysis} and relative cooling power
(RCP)~\cite{gschneidner_pecharsky2000magnetocaloric_materials}, defined as follows:

\begin{equation}
RC = \int_{T_1}^{T_2} |-\Delta S_{\mathrm{M}}(T)|\,\mathrm{d}T
\label{eqn:rc}
\end{equation}

\begin{equation}
RCP = -\Delta S_{\mathrm{M}}^{\mathrm{max}} \times \delta T_{\mathrm{FWHM}}
\label{eqn:rcp}
\end{equation}

\noindent where $T_{\mathrm{FWHM}}$ denotes full width at half-maximum of the $-\Delta S_{\mathrm{M}}(T)$ curve while
$T_1$ and $T_2$ refer respectively to the lower and higher limits of the $\delta T_{\mathrm{FWHM}}$ range.

The values of RC and RCP for the whole RE$_{11}$Co$_4$In$_9$ (RE = Gd--Er) family of compounds are listed in
Table~\ref{tbl:rcp_rc}. For a fixed value of the flux density change ($\Delta\mu_{0}H$), RC and RCP decrease
with the increasing number of the $4f$ electrons, reaching a local minimum for RE = Dy, and afterwards for 
RE = Ho, Er they increase to the values comparable with those found for RE = Gd.

The magnetocaloric performance of the RE$_{11}$Co$_4$In$_9$ (RE = rare earth) intermetallics is comparable to that of the best known
magnetocaloric materials~\cite{Gschneidner_et_al_2005_Recent_developments,Lyubina_2017_Magnetocaloric_materials}, making
RE$_{11}$Co$_4$In$_9$ good candidates for low-temperature magnetocaloric refrigeration.

\begin{table*}
\caption{The values of RCP and RC with $\Delta \mu_{0} H$ of 0--2, 0--5 and 0--9~T for RE$_{11}$Co$_4$In$_9$ (RE = Gd--Er).}
{
\begin{tabular*}{\textwidth}{@{\extracolsep{\fill}}llllllllll}
    \hline %
Material & \multicolumn{3}{c}{RCP [J$\cdot$kg$^{-1}$]} & \multicolumn{3}{c}{RC [J$\cdot$kg$^{-1}$]} & Ref.\\
\cline{2-4} \cline{5-7}
    & 0-2 T & 0-5 T & 0-9 T & 0-2 T & 0-5 T & 0-9 T \\ %
    \hline
    Gd$_{11}$Co$_4$In$_9$ & 106.3 & 357.9 &      & 81.5  & 269.9 &      & \cite{zhang2020structural}\\
    Tb$_{11}$Co$_4$In$_9$ & 44.6 & 179.5 & 522.1 & 34.5 & 139.5 & 391.2 & this work\\
    Dy$_{11}$Co$_4$In$_9$ & 27.1 & 128.4 &       & 20.4 & 97.8 &      & \cite{zhang2020structural}\\
    Ho$_{11}$Co$_4$In$_9$ & 87.4 & 306.7 &       & 66.1 & 228.7 &     & \cite{zhang2020structural}\\
    Er$_{11}$Co$_4$In$_9$ & 84.6 & 265.1 & 605.2 & 66.2 & 205.3 & 463.1 & this work\\
    \hline
\end{tabular*}}
\label{tbl:rcp_rc}
\end{table*}

\section{Conclusions}

The magnetocaloric effect in RE$_{11}$Co$_4$In$_9$ (RE = Tb, Er) has been investigated by magnetometric measurements
in the function of temperature and applied magnetic field. Based on these data, the following parameters characterizing
magnetocaloric performance have been determined: $-\Delta S_{\mathrm{M}}^{\mathrm{max}}$, TEC, RC and RCP. Comparison
of these parameters for the whole RE$_{11}$Co$_4$In$_9$ (RE = Gd--Er) family of compounds leads to the conclusion that the
intermetallics with RE = Gd, Ho and Er show high and comparable one to another magnetocaloric performance, much higher
than that found for RE = Tb and Dy. It is worth noting that although the parameters characterizing MCE in RE = Gd, Ho
and Er take similar values, the corresponding maximum magnetic entropy changes appear at different temperatures which
are closely related to the magnetic transition temperatures of individual compounds. The maximum of $-\Delta S_{\mathrm{M}}$
is found around 90, 20 and 10~K for RE = Gd, Ho and Er, respectively.

\section*{Declaration of Competing Interest}

The authors declare no conflict of interest.

\section*{Acknowledgements}

The research was partially carried out with the equipment purchased thanks to the financial support of the European
Regional Development Fund in the framework of the Polish Innovation Economy Operational Program (contract no.
POIG.02.01.00-12-023/08).

\bibliography{R11Co4In9_magn_effect.bib}

\end{document}